\documentclass[prb,nobalancelastpage,twocolumn,nofootinbib,superscriptaddress,showpacs]{revtex4-1}

\usepackage{color,amsthm,amsmath,amsfonts,graphicx,bm}
\usepackage{bm}
\usepackage{amsfonts}
\usepackage{amssymb}
\usepackage{amsmath}
\usepackage{epsfig}
\usepackage{graphicx}
\usepackage{enumerate}
\usepackage{multirow}
\usepackage{verbatim}
\usepackage{subfigure}
\usepackage{bbold}
\usepackage{soul}
\usepackage{scrextend}

\newcommand{\tr}{\mathrm{tr}}

\newcommand{\mr}[1]{\mathrm{#1}}

\newcounter{notes}
\stepcounter{notes}

\begin{document}

\title{Signatures of the Many-body Localized Regime in Two Dimensions}

\date{\today}

\author{Thorsten B. Wahl}
\affiliation{Rudolf Peierls Centre for Theoretical Physics, Clarendon Laboratory, Parks Road, Oxford OX1 3PU, United Kingdom}

\author{Arijeet Pal}
\affiliation{Rudolf Peierls Centre for Theoretical Physics, Clarendon Laboratory, Parks Road, Oxford OX1 3PU, United Kingdom}

\author{Steven H. Simon}
\affiliation{Rudolf Peierls Centre for Theoretical Physics, Clarendon Laboratory, Parks Road, Oxford OX1 3PU, United Kingdom}

\begin{abstract}
Lessons from Anderson localization highlight the importance of dimensionality of real space for localization due to disorder. More recently, studies of many-body localization have focussed on the phenomenon in one dimension using techniques of exact diagonalization and tensor networks. On the other hand, experiments in two dimensions have provided concrete results going beyond the previously numerically accessible limits while posing several challenging questions. We present the first large-scale numerical examination of a disordered Bose-Hubbard model in two dimensions realized in cold atoms, which shows entanglement based signatures of many-body localization. By generalizing a low-depth quantum circuit to two dimensions we approximate eigenstates in the experimental parameter regimes for large systems, which is beyond the scope of exact diagonalization. A careful analysis of the eigenstate entanglement structure provides an indication of the putative phase transition marked by a peak in the fluctuations of entanglement entropy in a parameter range consistent with experiments. 
\end{abstract}

\maketitle

Many-body localization (MBL) is a paradigm shift in out-of-equilibrium quantum matter. This novel phase of matter is characterized by the absence of thermalization~\cite{anderson1958absence, basko2006metal, gornyi2005interacting, pal2010mb, oganesyan2007localization, nandkishore2015many}. An MBL system retains a memory of its initial state and displays only a logarithmic growth of entanglement following quantum quenches~\cite{Bardason2012}.
By localizing the excitations, MBL can also protect certain forms of topological and symmetry-breaking orders in excited states and provides an opportunity to process quantum information in a system driven far from equilibrium~\cite{Huse2013LPQO, Pekker2014HG, bahri2015localization, Chandran2014SPT}. The quantum phase transition separating the MBL and thermal phases poses a major challenge for developing a theory of dynamical critical phenomena described by many-body entanglement in highly excited states~\cite{kjall2014many, VHA_MBLTransition, Potter:2015ab, Chandran:2015ab, Lim2016MPS, Khemani2017MBLT, Dumitrescu2017}. 

It is well-known that dimensionality of real space affects single particle Anderson localization where in one and two dimensions (without spin-orbit coupling and broken time-reversal symmetry), the entire spectrum of single particle eigenstates is localized for arbitrarily weak disorder~\cite{Lee_RMP1985}. Although the defining properties of MBL in one dimension are firmly established both theoretically and experimentally \cite{imbrie2016many, Schreiber842}, the existence of the phenomenon in two and higher dimensions is much debated~\cite{chandran2016higherD, Roeck2016stability, deRoeck2017Stability, Agarwal2017RareregionReview, luitz2017smallbaths, ponte2017thermalinclusions,stability_Altman}. Experiments in cold atoms measuring local and global equilibration have shown the persistence of quantum memory for long times providing indications of an MBL-like phase in two dimensions~\cite{Choi1547, bordia2017quasiperiodic2D,stability_Altman}. On the other hand, theoretical criteria suggest that the lifetime of local memory is finite, albeit extremely long~\cite{chandran2016higherD, Roeck2016stability}.   

In this article, we evaluate the eigenstates of bosons hopping in a disordered lattice in two dimensions with on-site interactions. We generalize the tensor network method developed earlier for one dimensional systems to approximate the eigenstates in the localized regime \cite{Wahl2017PRX}. 
Because the system sizes accessible by our method are much larger than those that can be currently achieved with prior methods, we are for the first time able to locate the transition to the thermal phase~\cite{Geraedts_PRB2016, Inglis_PRL2016, thomson2017Flow}. 
We study the model in parameter regimes directly relevant to the experiments. The approximate eigenstates evaluated by our method exhibit signatures of a phase with low entanglement. By studying the distribution of entanglement entropies of a single site, we probe the eigenstates in the localized regime at large disorder. On lowering the disorder strength, the distribution of entanglement becomes bimodal with increasing weight at larger entanglement, indicating the transition into the thermal phase. Furthermore, we evaluate the energy-resolved transition point, which gives rise to a mobility edge, similar to the thermal-to-MBL transition in one dimension~\cite{luitz2015many}. 
We find a critical disorder strength which is larger than the one measured in optical lattice experiments using relaxation of a half-filled harmonic trap~\cite{Choi1547}. However, more recent experiments initialized with a charge density wave (CDW)
suggest that the transition might be at a higher disorder strength than found in earlier work, and the extraction of the critical point from the experimental data is a subtle point~\cite{bordia2017quasiperiodic2D,MBL_heat_bath}.

The approximate nature of our method prevents us from drawing final conclusions on the existence of MBL in two dimensions. However, due to the extremely long lifetimes of local memories, strongly disordered two-dimensional  systems might be viewed as many-body localized for all experimental and technological purposes. Our method is able to capture this MBL-like behavior on the relevant time scales, as explained in the following.


\vspace{12pt}
\begin{figure*}
\begin{picture}(240,240)
\put(0,0){\includegraphics[width=0.49\textwidth]{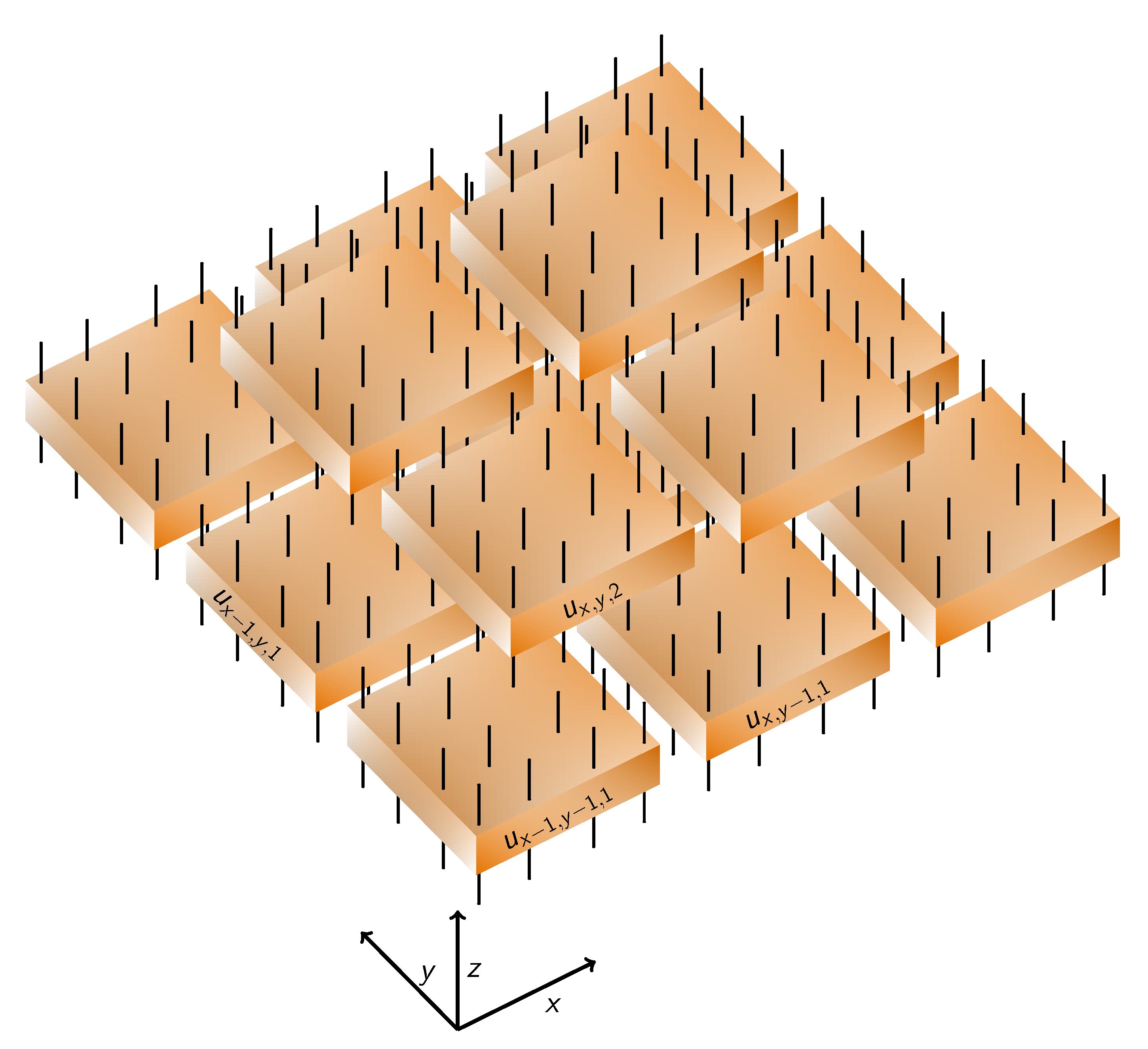}}
\put(0,230){\textbf{a}}
\end{picture} \ \
\begin{picture}(240,240)
\put(0,0){\includegraphics[width=0.49\textwidth]{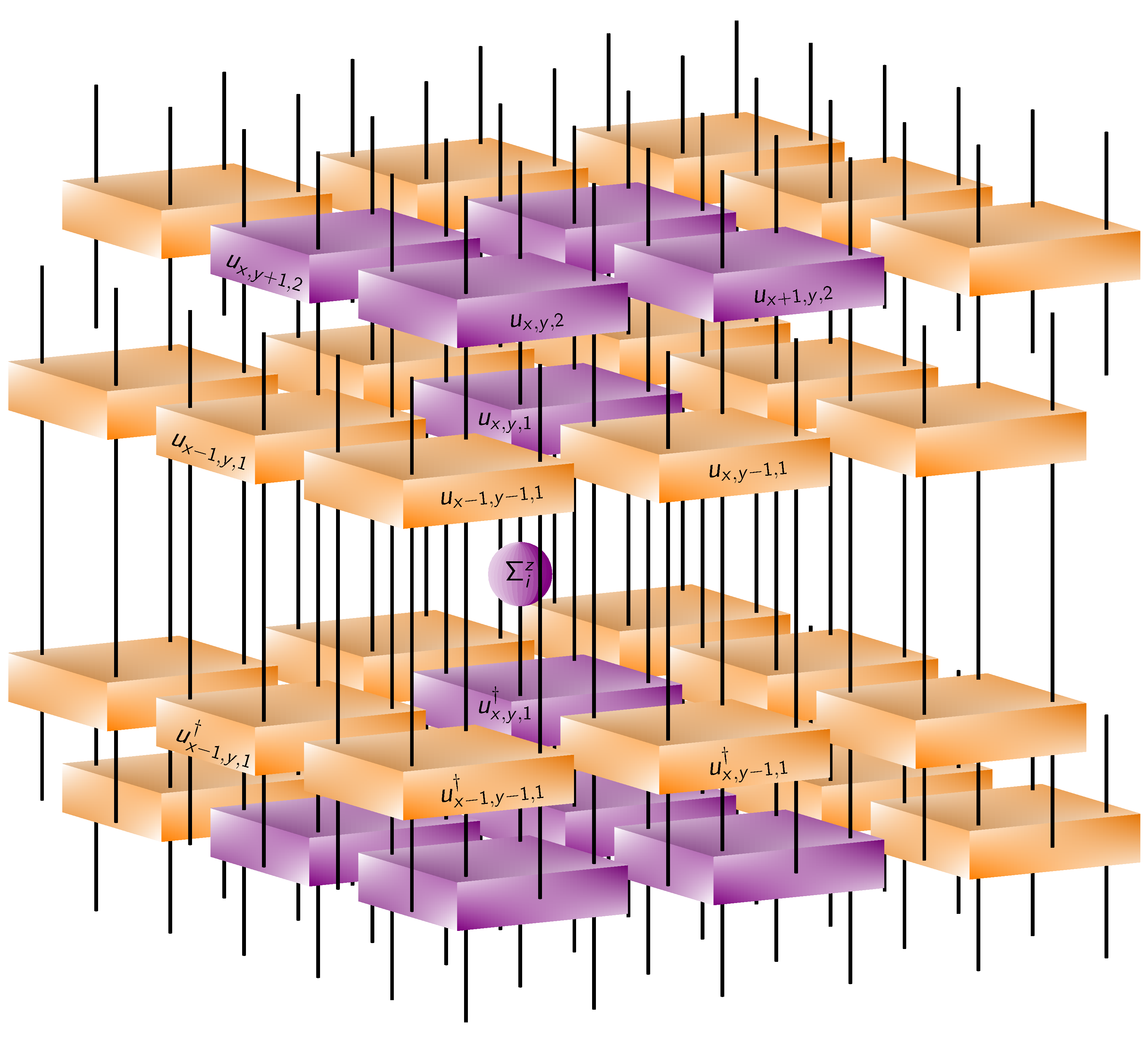}}
\put(0,230){\textbf{b}}
\end{picture}
  \caption{\textbf{a}, Tensor network $\tilde U$ composed of two layers of unitaries $u_{x,y,z}$ represented by orange boxes, which act on blocks of $\ell \times \ell$ sites (in the picture $\ell = 4$). The upper open legs correspond to the individual sites the unitaries act on, connected legs indicate tensor contractions and the lower open legs correspond to the approximate l-bit basis.
\textbf{b}, Evaluation of $\tilde \tau_i^z = \tilde U \Sigma_i^z \tilde U^\dag$. $\Sigma_i^z$ is represented by a violet sphere. All unitaries apart from the violet ones (causal cone) cancel, rendering $\tilde \tau_i^z$ non-trivial on a plaquette of size $2 \ell \times 2 \ell$. For simplicity, we depict $\ell = 2$.}
\label{fig:TNS}
\end{figure*}

\vspace{12pt}
\noindent\textbf{MBL in one and higher dimensions} \\
In one dimension, fully many-body localized systems are characterized by an extensive number of quasi-local operators, known as l-bits ($\tau_i^z$)~\cite{serbyn2013local, huse2014phenomenology, ros2015integrals, ImbrieLIOMreview2017}. These operators commute exactly with the Hamiltonian and also with each other. Therefore, all energy eigenstates are represented as a string of eigenvalues of the l-bits. Locality of the l-bits forces the eigenstates to satisfy the area law of entanglement \cite{Bauer:2013jw}. 
As a result, the unitary transformation ($U$) which diagonalizes the Hamiltonian can be efficiently approximated by a tensor network (TN) \cite{pekker2014encoding, Chandran2015STN, Pollmann2016TNS} where all eigenstates can be represented as matrix product states \cite{Friesdorf2015, yu2015finding, Khemani2016MPS}. Throughout this work, we consider shallow quantum circuits, which are TNs which can be contracted efficiently in any dimension. They are composed of  layers of local unitaries with each unitary acting on a finite contiguous block of spins~\cite{Wahl2017PRX}.  For the optimal set of local unitaries, the TN transforms the Hamiltonian into a predominantly diagonal form. 
This approach has been successfully implemented for large systems to 
estimate the location of the phase transition in one dimension \cite{Wahl2017PRX} and used analytically to prove the robustness of time reversal symmetry protected MBL phases~\cite{Wahl2017}. This method can also be used to construct the l-bits close to the MBL transition \cite{Kulshreshtha2017}.

\begin{figure*}[t]
\begin{picture}(200,200)
\put(0,0){\includegraphics[width=0.35\textwidth]{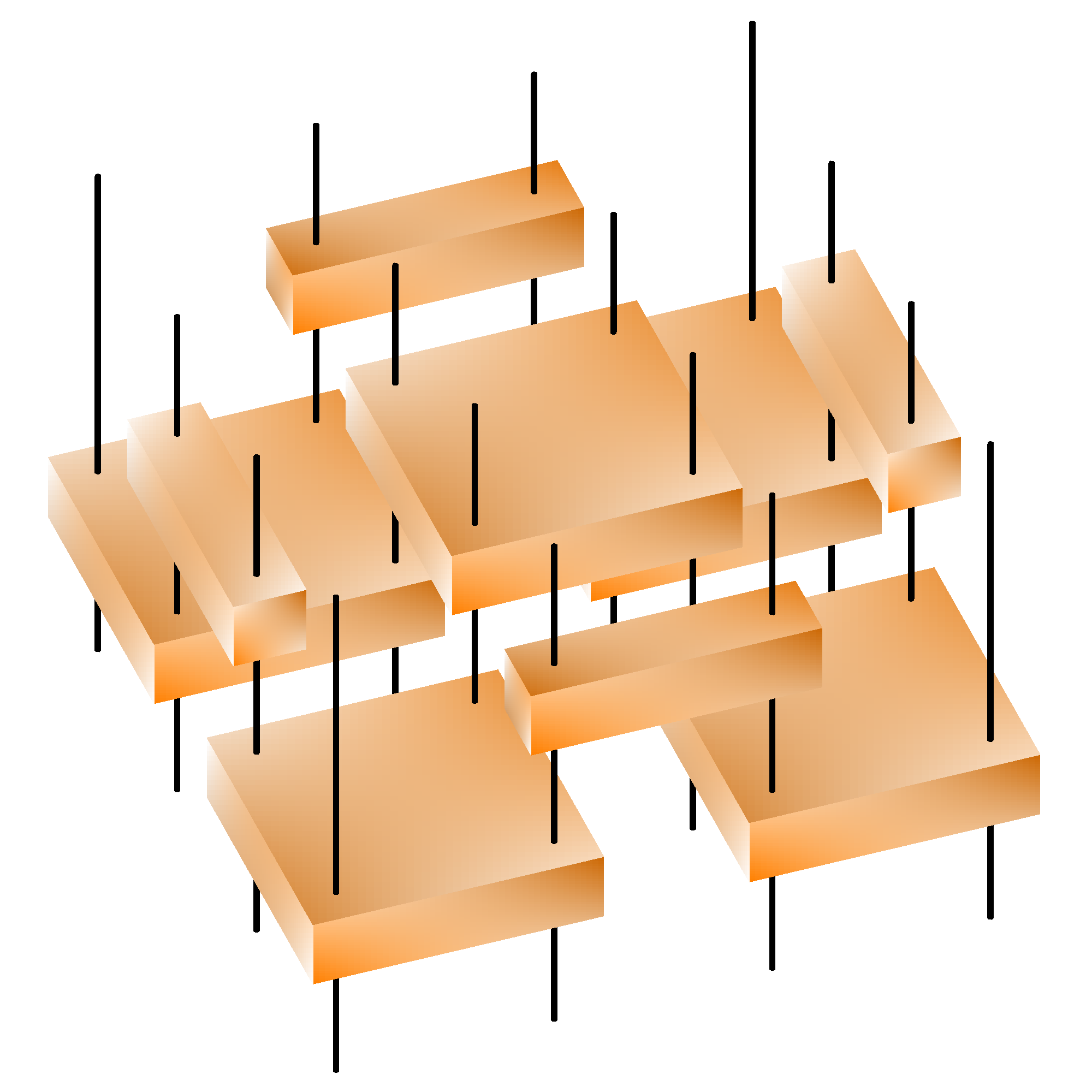}}
\put(-20,180){\textbf{a}}
\end{picture} \ \ \ \
\begin{picture}(200,200)
\put(0,0){\includegraphics[width=0.45\textwidth]{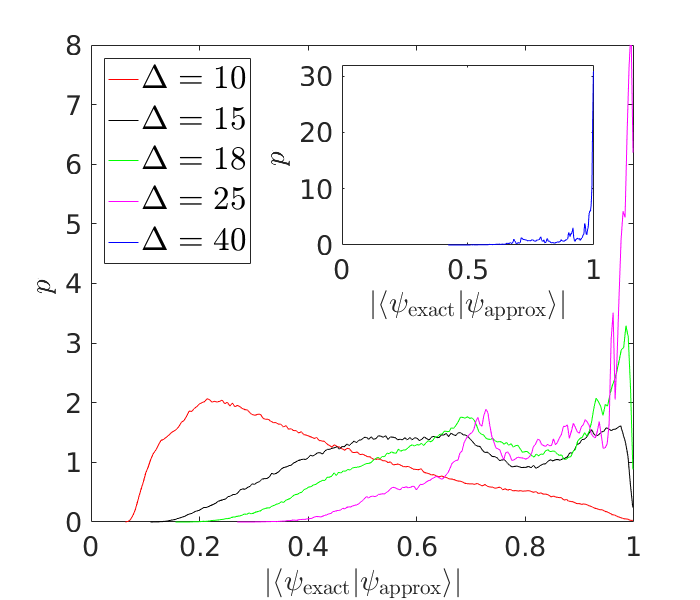}}
\put(-20,180){\textbf{b}}
\end{picture}
  \caption{\textbf{a}, Quantum circuit used to approximate the unitary diagonalizing the Hamiltonian on a $4 \times 4$ lattice with open boundary conditions. \textbf{b}, Probability density of the overlaps of the exact and matched approximate eigenstates. ($p$ is normalized such that it integrates to 1.) The inset shows the result for $\Delta = 40$.} 
\label{fig:ED}
\end{figure*}

In dimensions greater than one, the l-bit phenomenology can break down due to resonant interactions which destroy the exact conservation of the set of quasi-local operators~\cite{chandran2016higherD, Roeck2016stability, deRoeck2017Stability} $\tau_i^z$. As a result, for a finite-size system thermalization can be restored at late times. However, if local operators remain approximately conserved, the system displays features analogous to MBL on experimentally relevant time scales. Unlike in one dimension where the l-bits are exactly conserved due to the suppression of resonances even for a finite system size, in higher dimensions the proliferation of resonances is believed to be exponentially sensitive to perturbations. Therefore, MBL giving rise to a stable quantum phase may only be well-defined as the system size tends to infinity, although there are arguments in favor of instability of MBL even in this asymptotic limit~\cite{Roeck2016stability, deRoeck2017Stability}. We use the semantics of quantum phase transitions to describe the features in our finite-size calculations,  even though our numerical investigation may not capture this asymptotic limit. However, the error in our approximation provides a lower bound on the time scale over which local observables will decay in this system. 
Hence, while not answering the question of the existence of exact l-bits in two dimensions, we expect our approach to describe the experimentally observed localization phenomena.
Indeed, the conditions required to confirm the breakdown of conservation of these quasi-local operators likely require measurements for astronomically long times for a physical choice of parameters~\cite{Choi1547}. 


\vspace{12pt}
\noindent\textbf{Tensor network ansatz} \\
We propose a shallow TN capable of approximating the two-dimensional localized phase. It encodes a unitary $\tilde U$ which approximately diagonalizes the Hamiltonian $H$. We consider an $N \times N$ lattice with periodic boundary conditions. The TN is composed of two layers of smaller unitaries $u_{x,y,z}$ acting on $\ell \times \ell$ sites ($\ell$ even) as shown in Fig.~\ref{fig:TNS}a. It is a natural extension of the ansatz in Ref.~\onlinecite{Wahl2017PRX}. 
The approximately conserved operators $\tilde{\tau}_i^{z}$ are given by $\tilde \tau_i^z = \tilde U \Sigma_i^z \tilde U^\dag$ ($\Sigma_i^z$ is the spin-$z$ operator at site $i$). These operators commute mutually by construction. $\tilde{\tau}^z_i$ approximates $\tau_i^z$, which is itself expected to be only approximately conserved, but acts as a local memory on the experimentally relevant time scales. Hence, the TN can be optimized by minimizing the (squared) trace norm of the commutator with the Hamiltonian,
\begin{align}
f(\{u_{x,y,z}\}) &= \frac{1}{2^{N^2+1}} \sum_{i=1}^{N^2} \tr\left([H,\tilde \tau_i^z] [H,\tilde \tau_i^z]^\dag\right) \notag \\
&= \frac{1}{2^{N^2}} \sum_{i=1}^{N^2} \left(\tr(H^2 (\tilde \tau_i^z)^2) - \tr((\tilde \tau_i^z H)^2\right). \label{eq:fom_general}
\end{align}
$\tilde \tau_i^z$ corresponds to the TN contraction shown in Fig.~\ref{fig:TNS}b, which is non-trivial only in the region covered by the unitaries $u_{x,y,1}$, $u_{x,y,2}, u_{x+1,y,2}, u_{x,y+1,2}, u_{x+1,y+1,2}$.
As a result, for a nearest-neighbor Hamiltonian, Eq.~\eqref{eq:fom_general} decomposes into local parts,
\begin{align}
f&(\{u_{x,y,z}\}) = \mr{const.} \notag \\
&- \sum_{x,y=1}^{N/\ell} f_{x,y} (u_{x,y,1},u_{x,y,2},u_{x+1,y,2},u_{x,y+1,2},u_{x+1,y+1,2}). \label{eq:fom_local}
\end{align} 

The explicit representation of $f_{x,y}$ as a fully contracted TN can be found in Methods (Sec. I). 
In the general case, the length of the region within which $\tilde \tau_i^z$ is non-trivial is of order $\ell$. Therefore, in the MBL regime, the error in representing ${\tau}_i^{z}$ is expected to decrease exponentially with $\ell$. 

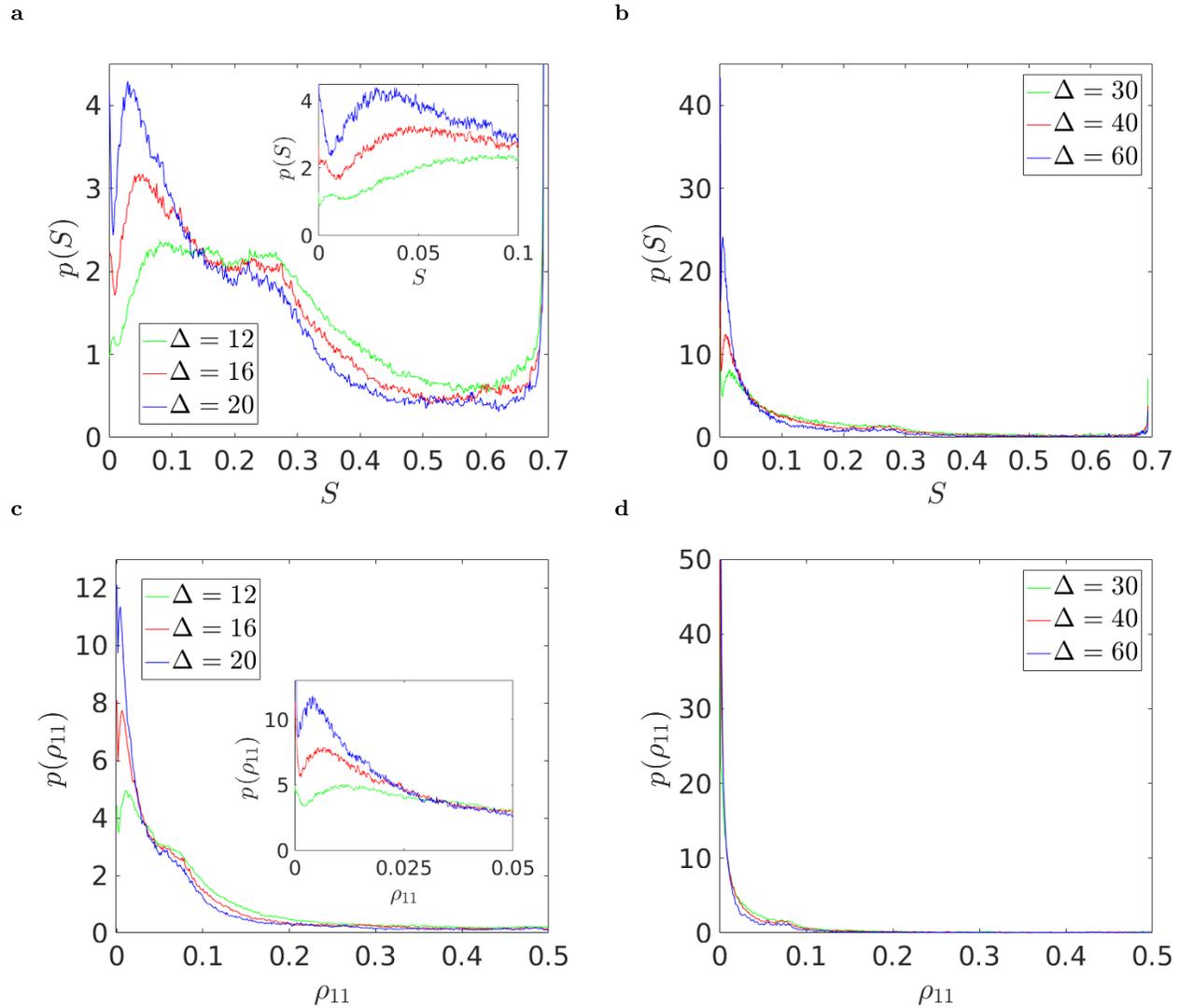
\begin{figure*}
\begin{picture}(220,200)
\put(0,0){\includegraphics[width=0.45\textwidth]{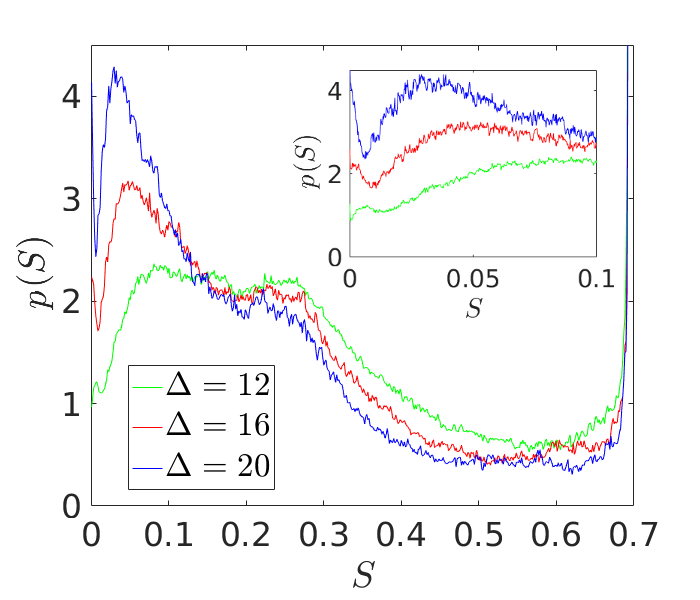}}
\put(-10,200){\textbf{a}}
\end{picture}
 \ \ \ \ \ \ \
\begin{picture}(220,200)
\put(0,0){\includegraphics[width=0.45\textwidth]{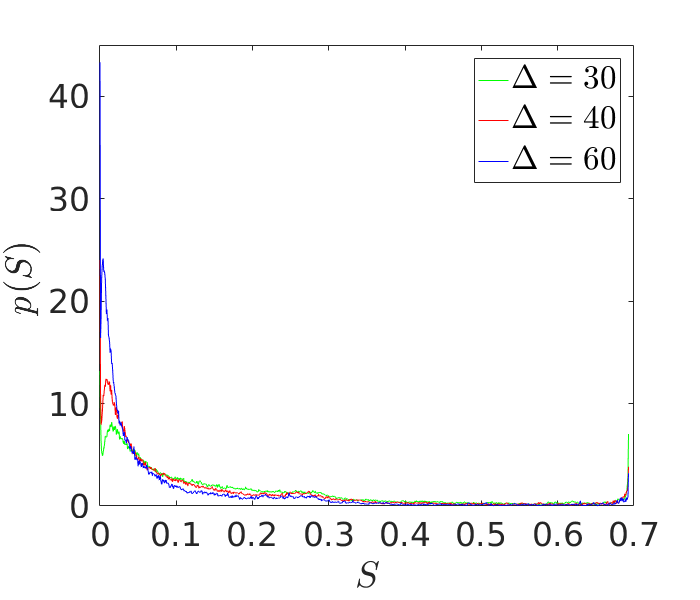}}
\put(-10,200){\textbf{b}}
\end{picture} 
\\
\begin{picture}(220,200)
\put(0,0){\includegraphics[width=0.45\textwidth]{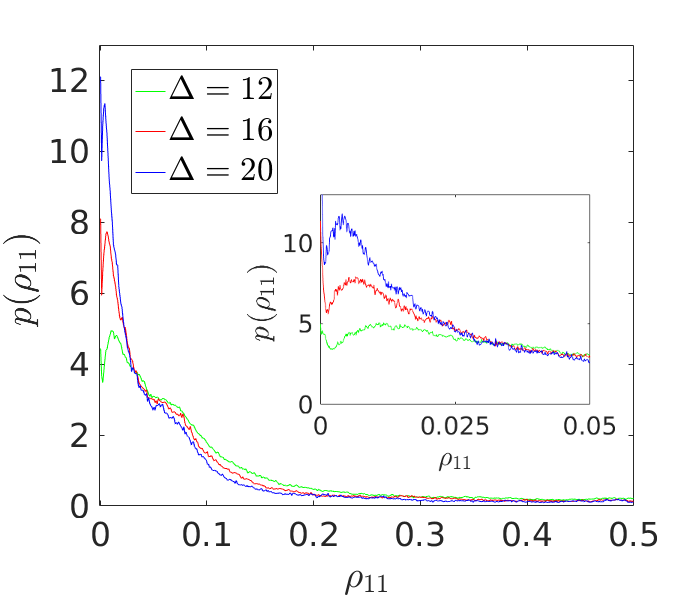}}
\put(-10,200){\textbf{c}}
\end{picture}
 \ \ \ \ \ \ \
\begin{picture}(220,200)
\put(0,0){\includegraphics[width=0.45\textwidth]{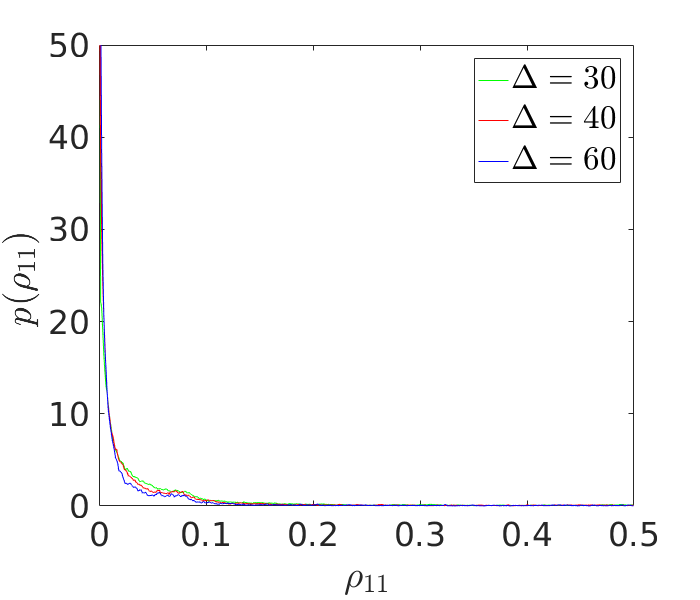}}
\put(-10,200){\textbf{d}}
\end{picture} 
  \caption{\textbf{a}, \textbf{b}: Probability density of the on-site entropies for all eigenstates, disorder realizations and site positions for $n_\mr{max} = 1$. \textbf{c}, \textbf{d}: Probability density of the corresponding on-site occupations. The distributions are only shown for $0\leq \rho_{11} \leq 0.5$, as they are symmetric around $\rho_{11} = 0.5$. The insets are enlargements to show the bimodal features more clearly.
\label{fig:entropies_exp}}
\end{figure*}

\vspace{12pt}
\noindent\textbf{Simulation of the experimental model}\\ 
In the following, we use our TN approach to investigate the MBL phase in a two-dimensional optical lattice~\cite{Choi1547,MBL_heat_bath}. The phase transition was determined experimentally 
by measuring the loss of memory of the initial conditions of bosonic Rubidium-87 atoms in a random optical potential. 
The dynamics of the atoms can be modeled by the disordered Bose-Hubbard Hamiltonian
\begin{align}
H = -J \sum_{\langle i, j \rangle} (a_i^\dag a_j +  a_j^\dag a_i) + \sum_{i} \left(\frac{U'}{2}  n_i (n_i - 1) + \delta_i n_i\right), \label{eq:exp_Ham}
\end{align}
where the nearest neighbor links $\langle i, j \rangle$ are counted once for each pair $(i,j)$, $a_i$ ($a_i^\dag$) are bosonic annihilation (creation) operators and $n_i = a_i^\dag a_i$ is the particle number operator. $\delta_i$ is a random variable chosen from a Gaussian distribution with full-width half-maximum $\Delta$, 
and we fix the energy scale through $J = 1$. Note that we neglect the trapping potential, as we are interested in the phase transition in the bulk of the system, i.e., the  center of the trap. 
We truncate the on-site occupation number to $n_\mr{max}$, which is an approximation suitable to describe the experiments where higher occupation number states are rarely encountered. (Experimentally, only up to 17\,\% of atoms are in doubly occupied sites~\cite{MBL_heat_bath}.)  We perform calculations for $n_\mr{max} = 1$ (corresponding to hard core bosons or spin-$\frac{1}{2}$ particles) and $n_\mr{max} = 2$ (spin-$1$).


In order to benchmark our method, we compared the optimized approximate eigenstates with the exact ones for a $4 \times 4$ system with open boundary conditions for $\ell = 2$ and $n_\mr{max} = 1$ using 30 disorder realizations for various $\Delta$, cf.~Fig.~\ref{fig:ED}. 
In the thermal phase, but close to the phase transition (which we will later determine to be at $\Delta_c \approx 19$), the overlap distribution has a double peak structure of poorly approximated eigenstates and well approximated eigenstates. This indicates the presence of both delocalized eigenstates, which our TN ansatz fails to capture, and localized eigenstates, which it approximates well even in the thermal phase. This is the underlying reason why we were able to obtain the mobility edge, as outlined below. In the localized phase ($\Delta > 20$), almost all approximate eigenstates have more than 50\,\% overlap with the exact ones.

\begin{figure}[t]
\includegraphics[width=0.5\textwidth]{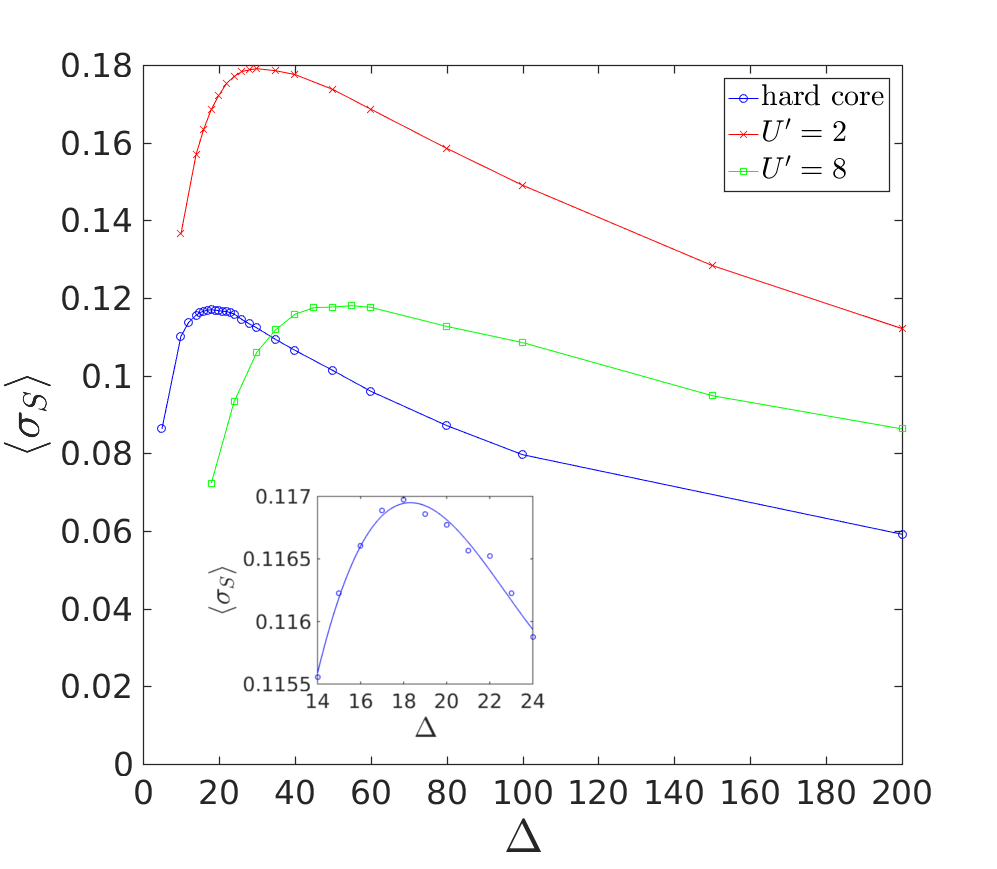}
  \caption{Variance of the entropy as a function of $\Delta$ for 30 disorder realizations. Blue circles correspond to the hard-core limit $n_\mr{max} = 1$ and the other data points to $n_\mr{max} = 2$. 
  For $n_\mr{max} = 1$, the entropies were averaged over all approximate  eigenstates and for $n_\mr{max} = 2$ over $10^4$ approximate eigenstates per lattice site (symbols are bigger than the sampling error). After evaluating the variance of the average on-site entropy with respect to disorder realizations, the average over sites was calculated to improve smoothness~\cite{Wahl2017PRX}. The inset shows an enlargement of the maximum for $n_\mr{max} = 1$ and a cubic fit, indicating a transition at $\Delta_c^{n_\mr{max} = 1} = 18.3$. Fits with simple rationals of polynomials yield similarly $\Delta_c^{U' = 2}  = 31$ and $\Delta_c^{U'=8} = 52$.
}    
\label{fig:S_exp}
\end{figure}

\vspace{12pt}

\noindent\textbf{Signatures of the MBL transition} \\
Our calculations were all carried out using $\ell = 2$ (which in one dimension gives already a fairly good approximation of the critical point of the disordered Heisenberg model~\cite{Wahl2017PRX}). 
A hallmark of the localized phase is the locality of entanglement even in the presence of interactions. A single site $i$ is only weakly coupled to its environment $\overline i$ and the reduced density matrix $\rho = \mr{Tr}_{\bar{i}} \left( |\psi_m \rangle \langle \psi_m | \right)$ of an eigenstate $|\psi_m\rangle$ 
remains close to a pure state. On the other hand, in the thermal phase, a single site is maximally entangled with its environment. 
We calculated the on-site entanglement entropies ($S$) for 30 disorder realizations on a $10 \times 10$ lattice ($n_\mr{max} = 1$) for various disorder strengths $\Delta$. Those entropies can be computed efficiently as explained in Methods (Sec. V). Their distributions are displayed in Fig.~\ref{fig:entropies_exp}a,b. For $\Delta \approx 20$, the distribution is bimodal with 
a sharp peak at $S=0$ along with a broad maximum at lower values of $S$. This provides evidence for the phase transition from the MBL to the thermal regime. The bimodality indicates the co-existence of highly localized states with low entanglement and delocalized states with larger entanglement in the critical regime. This resembles the behavior around the critical point of one-dimensional models exhibiting MBL~\cite{yu2016bimodal}. In contrast, for $\Delta \geq 40$, the distribution is mostly concentrated in a single peak at $S = 0$, where delocalized states are absent. Note that the peak at $\ln(2) \approx 0.69$ appears because of a singularity in the map from the on-site occupation $\rho_{11} = \langle n_i \rangle$ to the entropy $S$ at $\rho_{11} = 0.5$. 
The distribution of $\rho_{11}$ is shown in Fig.~\ref{fig:entropies_exp}c,d to provide data which can be compared to time-averaged atomic microscope measurements. It has no peak at $\rho_{11} = 0.5$.

We also computed the reduced density matrices $\rho_{2 \times 2}$ of all $2 \times 2$ plaquettes and extracted their entanglement spectra. They also display a double peak structure, which for $\Delta > 20$ is consistent with a localized phase (see Methods, Sec. V).

In order to locate the critical point more precisely, we evaluated the variance of the on-site entanglement entropies with respect to different disorder realizations shown in Fig.~\ref{fig:S_exp}:  
 For $n_\mr{max} = 1$, it peaks at $\Delta_c^{n_\mr{max}=1} = 18.3$.
 For $n_\mr{max} = 2$, the entropy fluctuations using 30 disorder realizations and $6 \times 6$ lattices indicate MBL-to-thermal transitions at $\Delta_c^{U' = 2} = 31$ and $\Delta_c^{U' = 8} = 52$, respectively. Note that the maxima get broader with increasing $U'$ due to the separation of energy scales between hopping and interaction terms, making it harder to track down the transition point numerically.  
 
 
\begin{figure*}[t]
\begin{picture}(220,200)
\put(0,0){\includegraphics[width=0.45\textwidth]{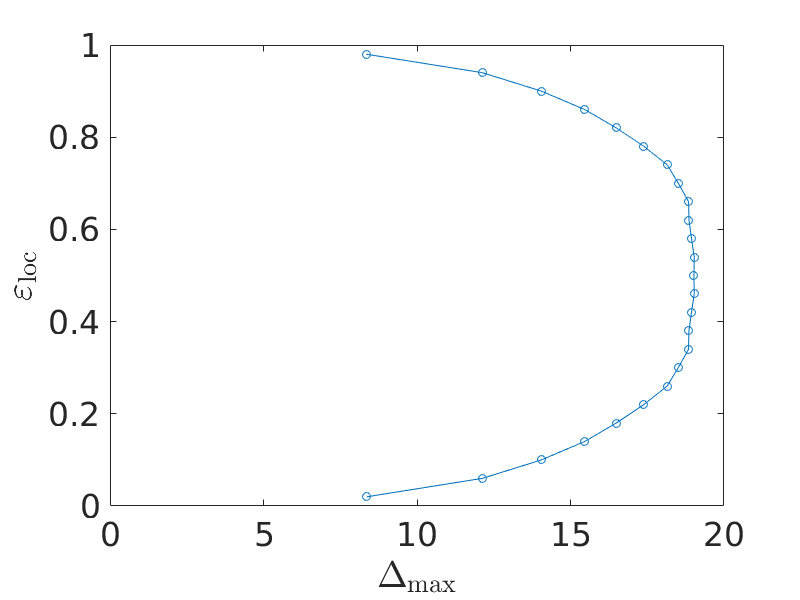}}
\put(-10,190){\textbf{a}}
\end{picture}
 \ \ \ \ \ \ \
\begin{picture}(220,200)
\put(0,0){\includegraphics[width=0.44\textwidth]{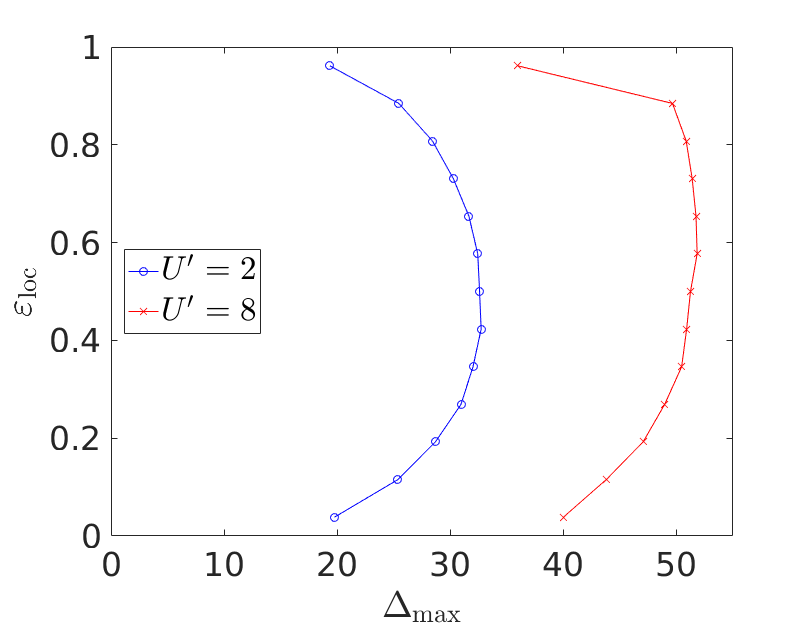}}
\put(-10,190){\textbf{b}}
\end{picture} 
  \caption{Mobility edge for $n_\mr{max} = 1$ (\textbf{a}) and $n_\mr{max} = 2$ (\textbf{b}) obtained as explained in the main text. 
  The extremal points of the mobility edges yield $\Delta_c^{n_\mr{max}=1} = 19.0$, $\Delta_c^{U'=2} = 33$ and $\Delta_c^{U'=8} = 52$.}    
\label{fig:mobility_edge}
\end{figure*}

Yet another compelling evidence for MBL-like behavior is the fact that the optimized quantum circuit displays a mobility edge, which we calculated as follows: The on-site entropies depend only on the (approximate) l-bit configuration in a $4 \times 4$ region around the corresponding site. 
Local energy contributions behave similarly. In contrast, by changing the l-bit indices outside the $4 \times 4$ region, the overall energy can be changed while keeping the on-site entropy invariant. We thus fixed those outer l-bits by averaging the total energy expectation value over all of them, giving rise to a local energy (up to an overall offset). For each $4 \times 4$ region, those local energies were rescaled such that they lie between $0$ and $1$, resulting in the quantity $\varepsilon_\mr{loc}$. This procedure is different from conventional ways of calculating the mobility edge, where the half-cut entanglement entropy is evaluated for certain total energy windows. For $n_\mr{max} = 1$ we evaluated the full set of l-bit configurations within the $4 \times 4$ regions and for $n_\mr{max} = 2$ we sampled over $10^4$ configurations for each one.
Based on that, we calculated the entropy fluctuation with respect to disorder realizations within certain energy windows $[\varepsilon_\mr{loc}, \varepsilon_\mr{loc} + d \varepsilon_\mr{loc}]$. The disorder strengths $\Delta_\mr{max}$ with maximal entropy fluctuations yield the mobility edges shown in Fig.~\ref{fig:mobility_edge}. 

When comparing to experimental results, one has to keep in mind that experiments so far were carried out at half filling (or less)~\cite{Choi1547,bordia2017quasiperiodic2D,MBL_heat_bath}. For the Hamiltonian whose local boson number is truncated to $n_\mr{max} \ge 2$, eigenstates with smaller filling have smaller energy. Experimentally, any number of bosons can be located at a given site, though such states are exponentially rare and will thus not be discernible in the dynamics. This effectively induces a maximal on-site occupation number. Hence, also in the experiment, lower fillings correspond to lower energies and, due to the mobility edge, to lower measured transition points~\cite{Choi1547}. 
Since our method yields an approximation of all eigenstates at once, our transition points are weighted averages over fillings. Those weights are given by a binomial distribution, i.e., the vast majority of eigenstates encoded in the unitary matrix $\tilde U$ have filling fraction close to ${n_\mr{max}}/{2}$. 
 Therefore, our predicted transition points refer to experiments carried out at filling fraction ${n_\mr{max}}/{2}$. 
Thus, we have to compare the experiments of Refs.~\onlinecite{Choi1547,MBL_heat_bath} to our $n_\mr{max} = 1$ results: 
Experimentally, $\Delta_c = 5.3$ was obtained first~\cite{Choi1547}, 
although the measured equilibration times may depend significantly on the wavelength of CDW initial states, which would thus affect the location of the observed critical point~\cite{bordia2017quasiperiodic2D,MBL_heat_bath}.  Our prediction of $\Delta_c \approx 19$ at half filling will possibly be borne out in experiments using initial CDW states with shorter wavelengths, where the relaxation is dominated by local equilibration. For such initial states, we expect the measured critical point to move to disorder strengths $\ge 50$ (for $U' \ge 8$) for filling fraction $1$.

\begin{figure}
\includegraphics[width=0.5\textwidth]{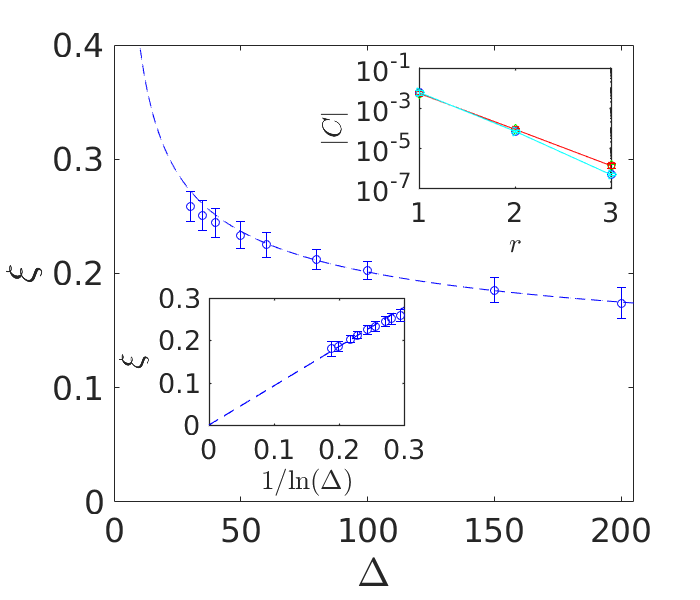}
  \caption{Dependence of the correlation length $\xi$ on the disorder strength $\Delta$ for $n_\mr{max} = 1$. The upper right inset shows the eigenstate (100 samples), site and disorder averaged correlation $C$ for $\Delta = 60$ as a function of distance $r$. The sites are separated along horizontal or vertical direction (indicated by different colors; error bars are smaller than the symbol size). 
The lower left inset displays the correlation length $\xi$ as a function of $1/\ln(\Delta)$. The dashed line represents a fit with $\xi \propto{1}/{\ln(\Delta)}$ to all data points with $\Delta \geq 60$.\label{fig:correlations}}
\end{figure}

Finally, for $n_\mr{max} = 1$ we calculated the density-density correlation function $C(r = |i-j|) =\langle n_i n_j \rangle - \langle n_i \rangle \langle n_j \rangle$, which is non-vanishing within a $6 \times 6$ region around site $i$ within the TN approximation.
The results for the correlation length $\xi$ in the strong disorder limit are shown in Fig~\ref{fig:correlations}. They indicate a slow decay, $\xi \approx a / \ln(\Delta)$ with $a = 0.93 \pm 0.07$, the same as in one dimension. This finding can be verified in optical lattice experiments with strong disorder and half filling. To provide further data which can be tested experimentally, we also calculated the transition point for random disorder taken from a uniform distribution $[-\Delta^\mr{uni},\Delta^\mr{uni}]$; we obtain $\Delta_c^{\mr{uni}} = 10.6$ for $n_\mr{max} = 1$, see Methods (Sec.~VI).

\vspace{24pt}
\noindent\textbf{Conclusions} \\
We have presented the first large-scale numerical study of eigenstates in the MBL regime of the disordered Bose-Hubbard model in two dimensions using shallow quantum circuits. 
 By characterizing the statistics of entanglement as a function of disorder, the location of the transition into the thermal phase has been estimated. The mobility edge displayed by our optimized tensor networks reflects the characteristics of a thermal-to-MBL transition, which further supports our conclusions.

Our work has important consequences for experiments with ultra-cold atoms studying quench dynamics in the presence of disorder. It provides an estimate of the MBL transition point in two dimensions for filling fraction 1 as a function of interaction strength, which can be verified in these experiments. We also presented quantitative predictions for the strongly disordered regime, which are expected to match experimental measurements to high accuracy. 
Large-scale simulations with higher $\ell$ will further elucidate the nature of the MBL regime in two dimensions.

\vspace{12pt}
\noindent\textbf{Acknowledgments} \\
We are grateful to Anushya Chandran and Chris Laumann for stimulating discussions and a careful reading of the manuscript, and to Antonio Abadal, Immanuel Bloch, Jae-Yoon Choi, and Christian Gross for detailed discussions related to the experiments. We would also like to thank David Huse, John Imbrie, Vadim Oganesyan, Wojciech De Roeck, Antonello Scardicchio, Shivaji Sondhi, and Thomas Spencer for fruitful discussions. S.H.S. and T.B.W. are both supported by TOPNES, EPSRC grant number EP/I031014/1. S.H.S. is also supported by EPSRC grant EP/N01930X/1. T.B.W. acknowledges usage of the University of Oxford Advanced Research Computing (ARC) facility in carrying out this work, http://dx.doi.org/10.5281/zenodo.22558. A.P. is supported by the Glasstone Fellowship and would like to thank the Aspen Center for Physics, which is supported by National Science Foundation grant PHY-1607611, and Simons Center for Geometry and Physics, Stonybrook University for their hospitality, where part of this work was performed. This project has received funding from the European Union’s Horizon 2020 research and innovation programme under the Marie Skłodowska-Curie grant agreement No. 749150. The contents of this article reflect only the authors' views and not the views of the European Commission. Statement of compliance with EPSRC policy framework on research data: This publication is theoretical work that does not require supporting research data.

\noindent\textbf{Author contributions:} \\
T.B.W. performed all the numerical simulations. The theoretical analysis and writing of the manuscript were jointly performed by  A. P., T B. W., and S.H.S.

\noindent\textbf{Corresponding author :} Arijeet Pal

\appendix

\begin{figure}
\includegraphics[width=0.37\textwidth]{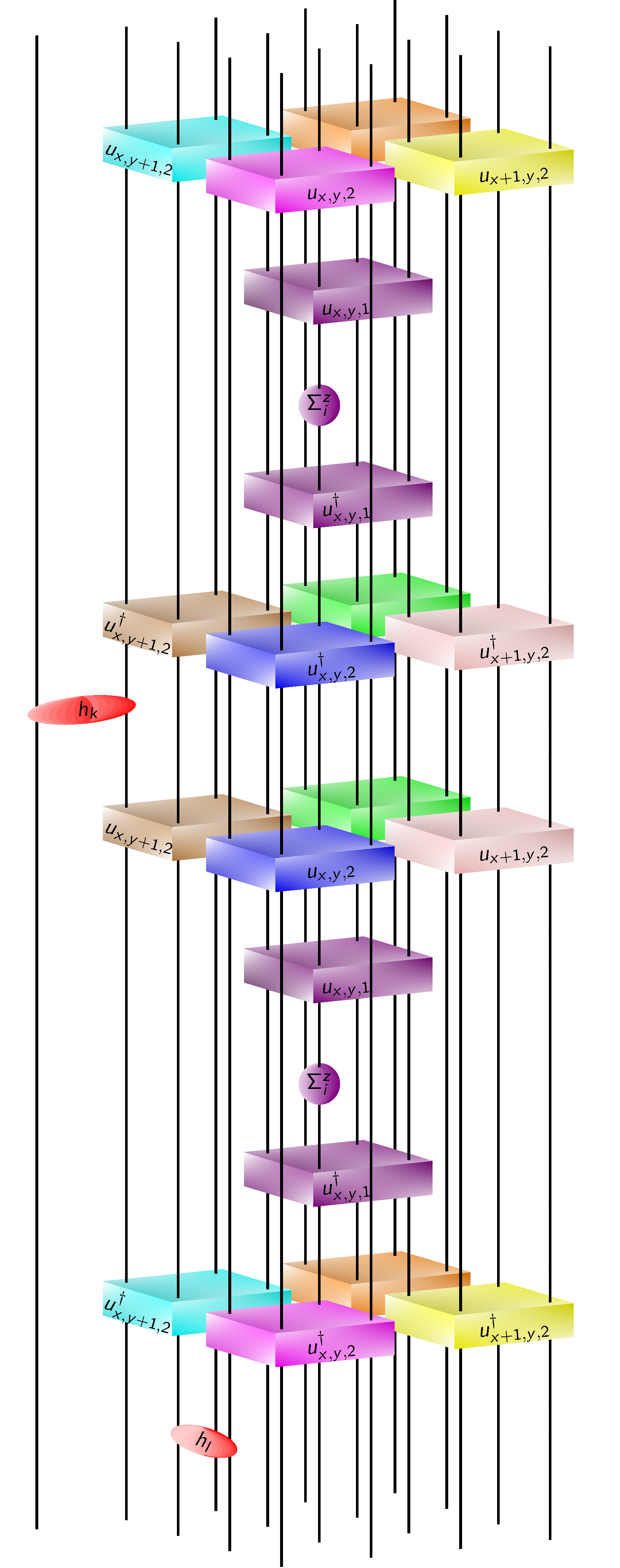}
\caption{Characteristic example of the tensor network contraction  $\mr{tr}\left(\tilde \tau_i^z h_k \tilde \tau_i^z h_l\right)$ yielding a non-trivial contribution to Eq.~\eqref{eq:fom_part2}, i.e., a contribution depending on the unitaries $\{u_{x,y,z}\}$. The figure is obtained by sandwiching two tensor networks $\tilde \tau_i^z$ as shown in Fig.~1b with the nearest neighbor terms $h_k$ (red ellipses). Due to the trace, the lines' upper and lower ends have to be pairwise contracted. Disconnected vertical lines have been ignored. After taking the trace, each of them leads to an overall prefactor of 2. For the computationally most efficient contraction, one first combines unitaries of the same color and includes $h_k$ and $\Sigma_i^z$ connected to them, leading to Fig.~\ref{fig:compact}. In the specific example in the picture, this is achieved by tracing out the left index of the upper Hamiltonian term and by splitting the lower Hamiltonian term into two on-site terms $h_L$, $h_R$ using a singular value decomposition, $h_l = \sum_{\mu} h_L^{(\mu)} h_R^{(\mu)}$, and summing over the tensor networks indexed by $\mu$. The strategy of contracting the tensor network is thus the same as in Ref.~\onlinecite{Wahl2017PRX}.}   
\label{fig:fom}
\end{figure} 
 
\section{Decomposing the commutator norm into local parts}

Here, we describe the contraction scheme to evaluate the commutator norm as defined in Eq. (1). We assume that the Hamiltonian is a sum of nearest neighbor terms, $H = \sum_k h_k$. In that case, the two terms in the final expression of Eq.~(1) are
\begin{align}
\tr\left(H^2 (\tilde \tau_i^z)^2\right) = \sum_{k,l} \tr \left(\tilde U (\Sigma_i^z)^2 \tilde U^\dag h_k h_l \right). \label{eq:fom_part1}
\end{align} 
and
\begin{align}
\tr\left((\tilde \tau_i^z H)^2\right) = \sum_{k,l} \tr \left(\tilde U \Sigma_i^z \tilde U^\dag h_k \tilde U \Sigma_i^z \tilde U^\dag h_l\right), \label{eq:fom_part2}
\end{align}
respectively. Using the representation for $\tilde \tau_i^z$ shown in Fig. 1b (which is non-trivial only in the region covered by the unitaries $u_{x,y,1}$, $u_{x,y,2}, u_{x+1,y,2}, u_{x,y+1,2}, u_{x+1,y+1,2}$), the right side of Eq.~\eqref{eq:fom_part2} can be evaluated as shown in Fig.~\ref{fig:fom}. The right side of Eq.~\eqref{eq:fom_part1} can be written as a similar tensor network contraction (this is most obvious if one inserts $\tilde U \tilde U^\dag$ between $h_k$ and $h_l$ leading to an expression which is formally equivalent to Eq.~\eqref{eq:fom_part2}). Therefore, $f(\{u_{x,y,z}\})$ can be expressed as in Eq.~(2). 

\begin{figure}
\includegraphics[width=0.36\textwidth]{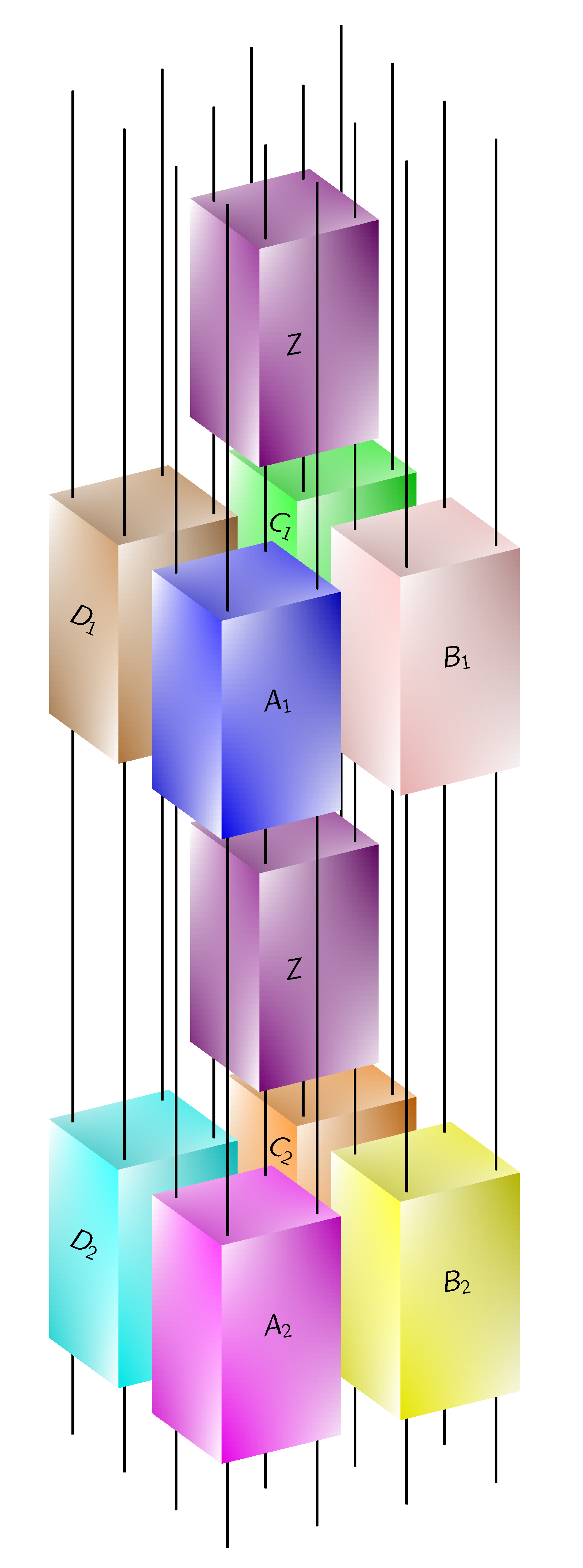}
\caption{Tensor network obtained after blocking matrices of the same color in Fig.~\ref{fig:fom} together (note that most of those new tensors will be identities). The lines' upper and lower ends have to be thought of as connected. One way of contracting the tensor network is by first contracting the tensors $A_1$ and $A_2$, $B_1$ and $B_2$, $C_1$ and $C_2$,  $D_1$ and $D_2$, resulting in new tensors $A_{12}, B_{12}, C_{12}, D_{12}$, respectively. The involved computational cost is of order $\left(2^{ \ell^2/2}\right)^2 \cdot 2^{3\ell^2/2} = 2^{5 \ell^2/2}$. Thereafter, one combines adjacent tensors (e.g., $A_{12}$ and $B_{12}$, $C_{12}$ and $D_{12}$) and contracts one of the resulting tensors with the lower $Z$ and the other tensor with the upper $Z$. Afterwards, the two obtained tensors are contracted, giving the desired overall contraction result. All those contractions have a computational cost of order $\left(2^{\ell^2}\right)^2 2^{\ell^2} = 2^{3 \ell^2}$.}    
\label{fig:compact}
\end{figure}

The tensor network contraction of Fig.~\ref{fig:fom} is carried out most efficiently as follows. First, one combines unitaries and their adjoints (shown in the same color in Fig.~\ref{fig:fom}) to new tensors. If there is no other tensor between them, this obviously results in the identity. In contrast, $u_{x,y,1}$ will be combined with $u_{x,y,1}^\dag$ and $\Sigma_i^z$ to form a new tensor $Z$. The other tensors might have to be combined with the Hamiltonian terms $h_k$ and $h_l$, respectively. The resulting tensor network is shown in Fig.~\ref{fig:compact}. A special case is the one of $k = l$ with the two Hamiltonian terms being connected by a separate line if they do not lie entirely within the region covered by the unitaries $u_{x,y,2}, u_{x+1,y,2}, u_{x,y+1,2}, u_{x+1,y+1,2}$. This squares the dimension of the index contraction corresponding to one of the lines in Fig.~\ref{fig:compact}. The leading contribution to the computational cost stems from the multiplication of $2^{\ell^2} \times 2^{\ell^2}$ matrices, i.e., it scales as $2^{3 \ell^2}$. As the number of tensor network contractions of the type shown in Fig.~\ref{fig:compact} is independent of the system size, each term $f_{x,y}$ 
 contributes  $2^{3 \ell^2}$ to the overall computational cost, which is thus of order $N^2 2^{3 \ell^2}$.

The derivative of our figure of merit can be calculated easily by cutting out the unitary matrix $u_{x,y,z}$ with respect to which the derivative is taken. Afterwards, one contracts the tensor network of Figs.~\ref{fig:fom} and~\ref{fig:compact} in such an order that the removed unitary would come last, cf. Ref.~\onlinecite{Wahl2017PRX}. This comes at a subleading increase of computational cost.

\vspace{12pt}
\section{Optimization method} 
We take advantage of the symmetries of the considered model in order to reduce the number of parameters: The Hamiltonian is real and possesses $U(1)$ symmetry, i.e., particle number conservation in the experimental model Eq.~(3) and spin-$z$ conservation in its spin representation, $[H, \sum_i \Sigma_i^z] = 0$. In general, such global symmetries can be imposed on the individual tensors which make up the tensor network without much loss of accuracy~\cite{Sanz2009,perez2010}. We do so by taking real unitary matrices $u_{x,y,z}$, i.e., orthogonal matrices, which on top of that preserve the spin-$z$ component individually, $[u_{x,y,z},\sum_{i \in \Box_{\ell \times \ell}} \Sigma_i^z] = 0$. Both taken together imply that the unitaries must have a block diagonal form (in the spin-$z$ basis),
\begin{align}
u_{x,y,z} = \bigoplus_{b}^{} e^{A_{x,y,z}^{(b)}},
\end{align}
where $A_{x,y,z}^{(b)} = -A_{x,y,z}^{(b)\top}$, a skew-symmetric (real) matrix. In each minimization run, the unitaries are initialized as identities and optimized using a quasi-Newtonian routine supplied by the derivative with respect to the parameters. We optimize one unitary at a time and sweep alternately over all the columns and all the rows of the $N \times N$ system (starting each sweep at site $(1,1)$). 

We verified numerically that no significant loss of accuracy occurs compared to a full parameterization of the unitaries.

%
%
%
\begin{figure}
\begin{picture}(220,200)
\put(0,0){\includegraphics[width=0.47\textwidth]{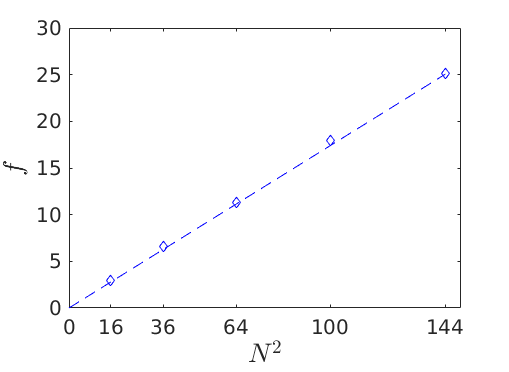}}
\put(0,180){\textbf{a}}
\end{picture}
 \\
\begin{picture}(220,200) 
 \put(0,0){\includegraphics[width=0.47\textwidth]{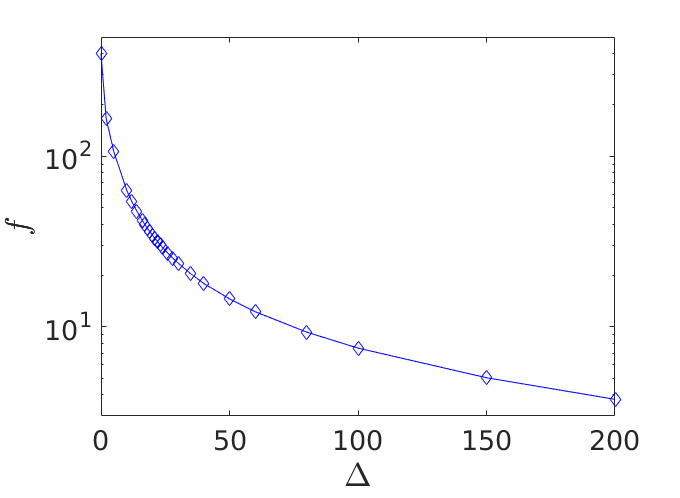}}
 \put(0,180){\textbf{b}}
\end{picture}
  \caption{\textbf{a}, Average optimized figure of merit as a function of system size $N^2$ for $\ell = 2$, disorder strength $\Delta = 40$ and 30 disorder realizations for each size. The dashed line indicates a linear fit. The errors of the mean are smaller than the symbol size. \textbf{b}, Average figure of merit as a function of disorder strength $\Delta$ for a $10 \times 10$ lattice using $\ell = 2$ and 30 disorder realizations. The errors of the mean are smaller than the symbol size.}    
\label{fig:size}
\end{figure}

\begin{figure*}
\begin{picture}(220,200)
\put(0,0){\includegraphics[width=0.45\textwidth]{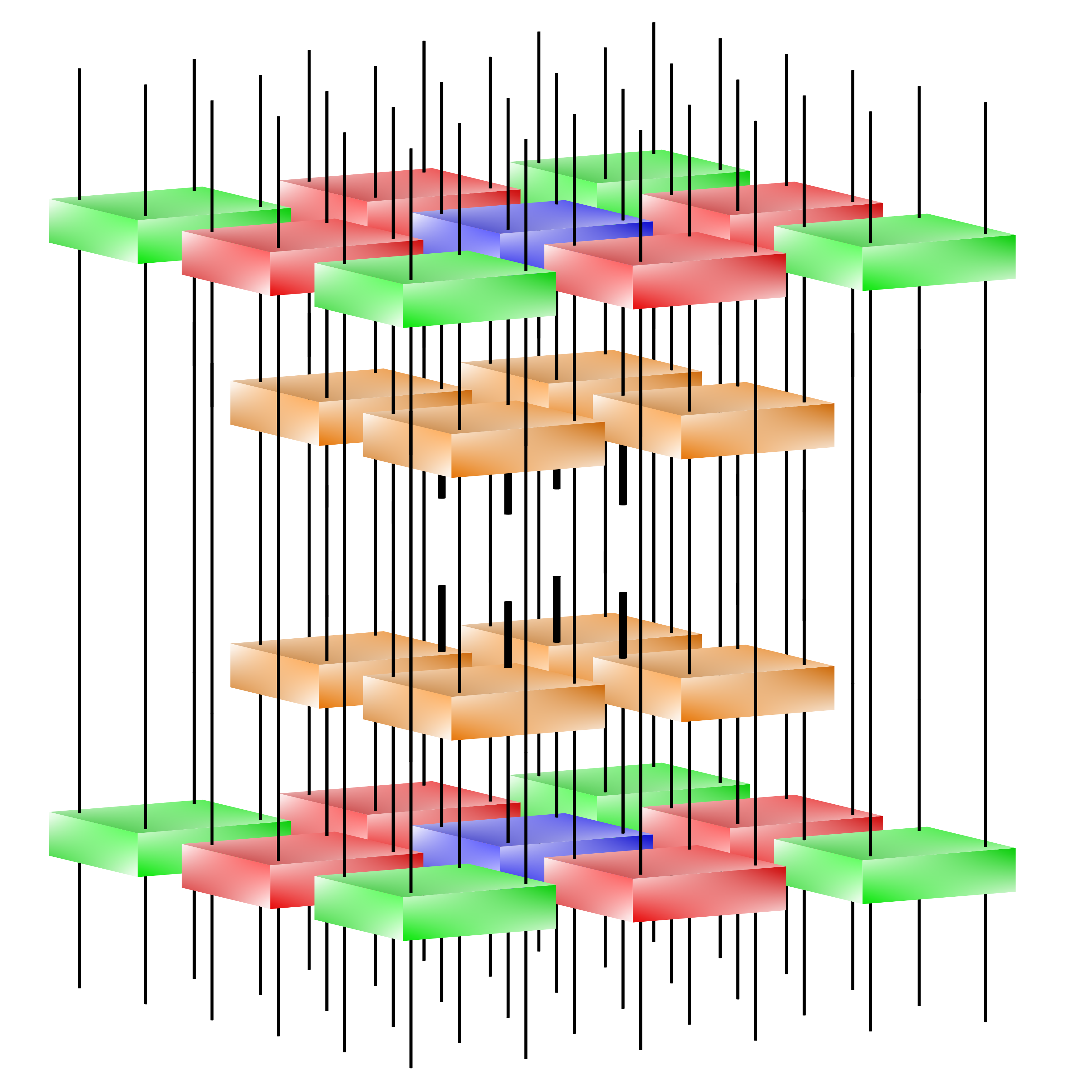}}
\put(-10,200){\textbf{a}}
\end{picture}
\ \ \ \ \ \ \ \ \
\begin{picture}(220,280)
\put(0,20){\includegraphics[width=0.45\textwidth]{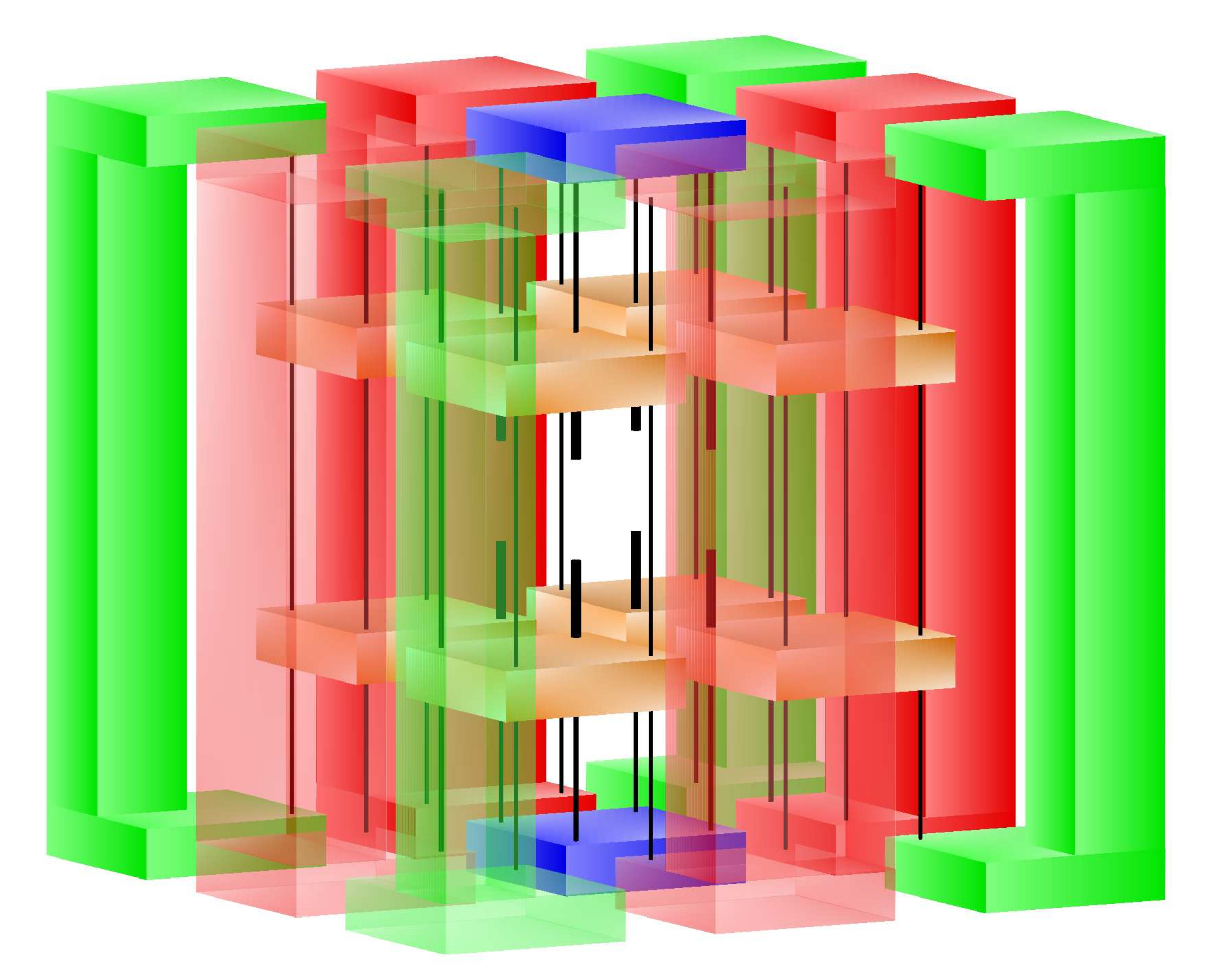}}
\put(-10,200){\textbf{b}}
\end{picture}
\caption{Calculation of reduced density matrices of $2 \times 2$ plaquettes for $\ell = 2$. We show the computationally hardest case, where the $2 \times 2$ region connects four unitaries of the upper layer of the tensor network. \textbf{a}, The indices of the lower and upper open legs are identical and given by the approximate l-bit configuration of the specific state under consideration. The thick open legs in the middle are the indices of the reduced density matrix. We first contract the red and green unitaries of the lower layer with their counterparts in the upper layer, resulting in the tensor network shown in \textbf{b}. After that, we contract every green tensor with the respective orange unitaries it is connected with. Then, we contract each of the resulting tensors with one of the red tensors leaving four big tensors. Thereafter, one contracts two pairs of such big tensors and the resulting two tensors first with the blue tensors and finally the new big two tensors with each other. The last step of the contraction is most expensive, corresponding to the multiplication of a $d^4 \times d^8$ matrix with a $d^8 \times d^4$ matrix (computational time $\sim d^{16}$).
}    
\label{fig:entropies}
\end{figure*}

\vspace{12pt} 
\section{Increasing the system size}
As the figure of merit decomposes into a sum of local parts, the expectation is it increases linearly with the system size on average. We confirmed this by studying 30 disorder realizations at $\Delta = 40$ for $n_\mr{max} = 1$ as a function of system size for periodic boundary conditions, see Fig.~\ref{fig:size}a.

In the MBL regime, eigenstates can be described by effective spin degrees of freedom~\cite{chandran2016higherD} $\tau_i^{z}$, which have exponentially decaying non-trivial matrix elements as a function of distance and commute approximately with the Hamiltonian. 
Therefore, for sufficiently large systems, we gain the following picture: All effective spin degrees of freedom which are located at a distance $r$ from a given region contribute to local properties of that region by an amount of order $2^{2 \pi r} e^{-r/\xi_L}$. Hence, their contribution vanishes exponentially with distance $r$ and the error made by approximating the eigenstates in terms of strictly short-range tensor network $l$-bits is independent of the system size. In other words, local observables can be approximated with a constant accuracy using our TN if the system size is increased. Note that at the same time the computational cost grows linearly with system size.


\vspace{12pt}
\section{Accuracy as a function of disorder strength} 
We also investigated the accuracy of our approximation as a function of disorder strength $\Delta$  for 30 disorder realizations. For different $\Delta$, only the overall strength of the random magnetic fields was adjusted to enhance comparability. The optimized figure of merit is shown in Fig.~\ref{fig:size}b. We find a strong, monotonic increase of the error with decreasing $\Delta$, especially below $\Delta = 50$. Notwithstanding, a concrete picture of the phase transition can only be gained from the distribution of entanglement entropies.

\begin{figure*}
\begin{picture}(220,200)
\put(0,0){\includegraphics[width=0.45\textwidth]{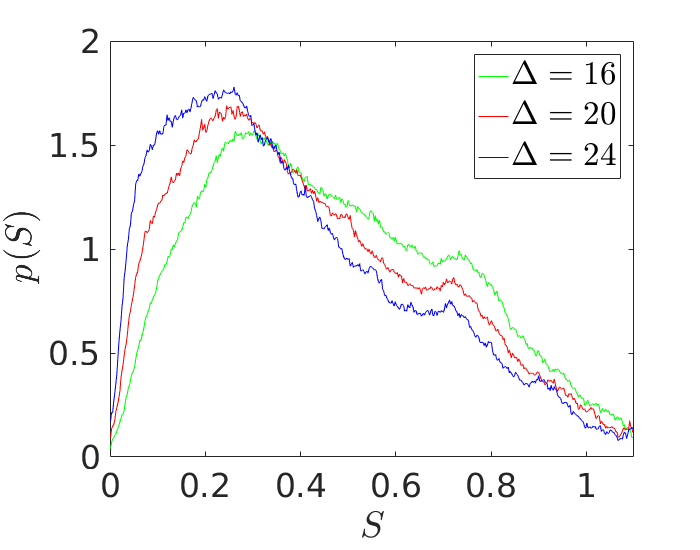}}
\put(-10,200){\textbf{a}}
\end{picture}
 \ \ \ \ \ \ \
\begin{picture}(220,200)
\put(0,0){\includegraphics[width=0.45\textwidth]{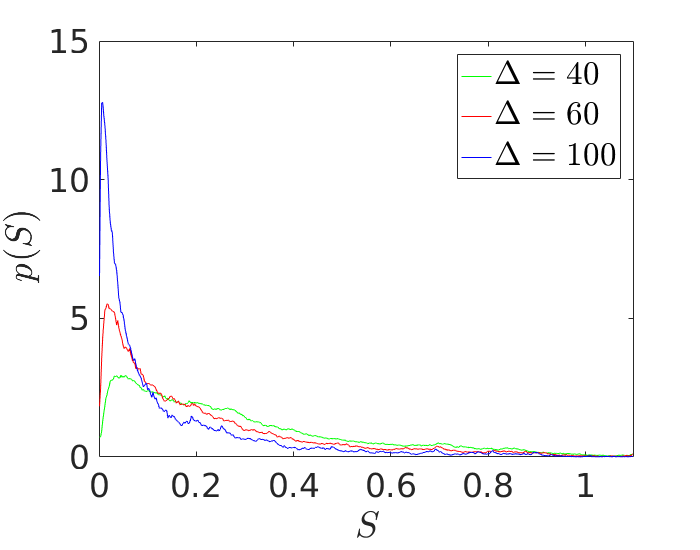}}
\put(-10,200){\textbf{b}} 
\end{picture} \\
\begin{picture}(220,200)
\put(0,0){\includegraphics[width=0.45\textwidth]{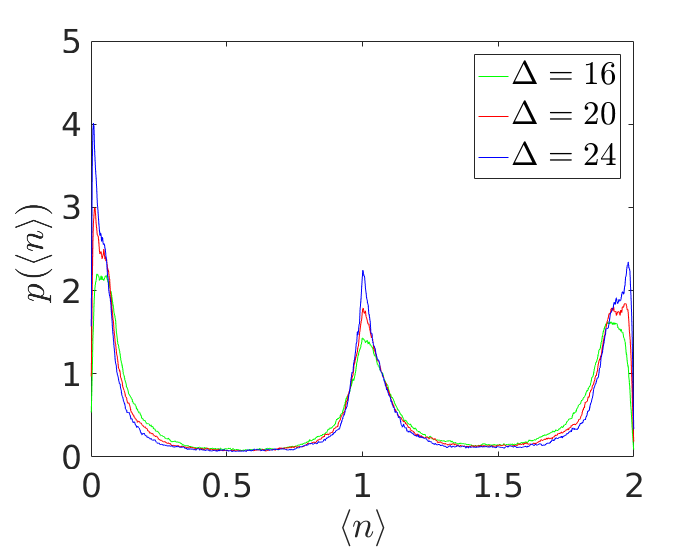}}
\put(-10,200){\textbf{c}}
\end{picture}
 \ \ \ \ \ \ \
\begin{picture}(220,200)
\put(0,0){\includegraphics[width=0.45\textwidth]{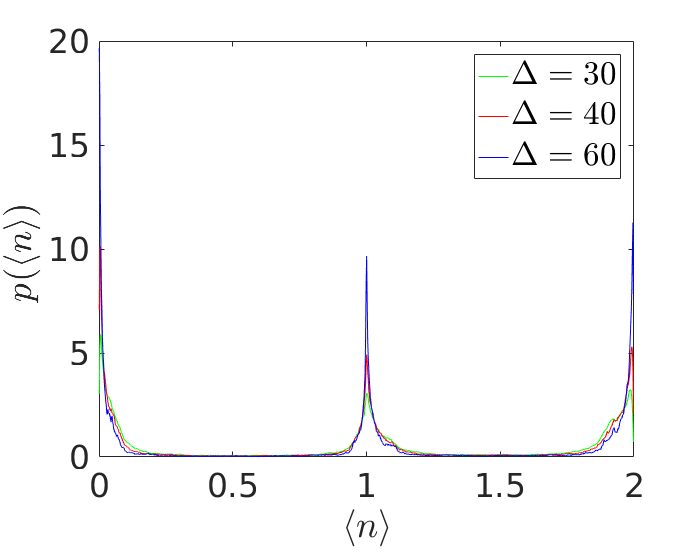}}
\put(-10,200){\textbf{d}}
\end{picture}
\caption{\textbf{a}, \textbf{b}, Probability density of on-site entanglement entropies for $U' = 2$, $n_\mr{max} = 1$ and 30 disorder realizations. \textbf{c}, \textbf{d}, Probability density of the corresponding on-site occupation $\langle n \rangle = \rho_{11} + 2 \rho_{22}$ (where $\rho$ is the $3 \times 3$ on-site reduced density matrix). The only discernable feature is that with increasing $\Delta$, the occupation expectation values tend to integer values, i.e., the sites increasingly disentangle from each other.\label{fig:entropies_U2}  } 
\end{figure*}

\vspace{12pt}
\section{Calculation of the entanglement entropies and spectra of $2 \times 2$ blocks of sites} 

In the following, we explain how the reduced density matrices of $2 \times 2$ plaquettes of approximate eigenstates given by our tensor network can be calculated efficiently. The computation of single-site reduced density matrices is a special case thereof. We will focus on the case of $2 \times 2$ plaquettes connecting four unitaries of the upper layer, which is the hardest one to evaluate. Other positions of the $2 \times 2$ blocks can be treated in a simplified way.

A $2 \times 2$ reduced density matrix of a given approximate eigenstate can be graphically represented by taking the tensor network in Fig. 1a, fixing its lower indices according to the  $l$-bit configuration of that approximate eigenstate and contracting it with its adjoint from above, leaving only the legs in the chosen $2 \times 2$ region open. After cancelling unitaries with their Hermitian conjugates, one obtains the tensor network contraction shown in Fig.~\ref{fig:entropies}a (for $\ell = 2$). Even though this tensor network is non-trivial in a Hilbert space of dimension $d^{36}$ ($d:$ Hilbert space dimension of a single site), its tensors can be contracted in such an order that the computational time is only of order $d^{16}$, cf. Fig.~\ref{fig:entropies}b.

\begin{figure*}
\begin{picture}(220,200)
\put(0,0){\includegraphics[width=0.45\textwidth]{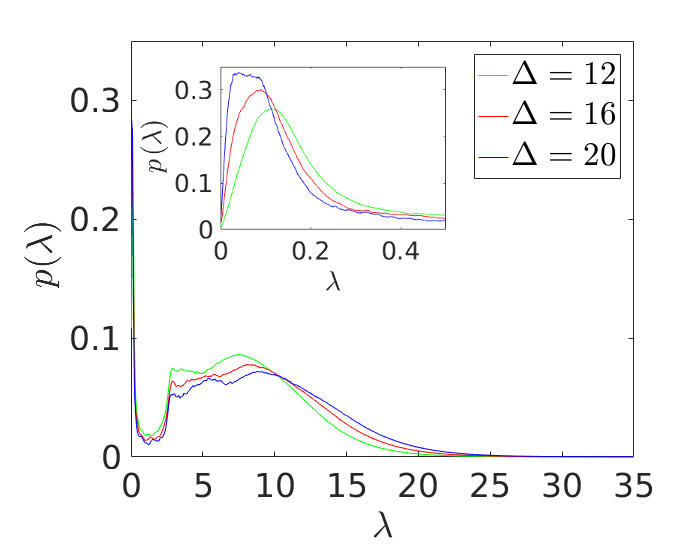}}
\put(-10,200){\textbf{a}}
\end{picture}
 \ \ \ \ \ 
\begin{picture}(220,200)
\put(0,0){\includegraphics[width=0.45\textwidth]{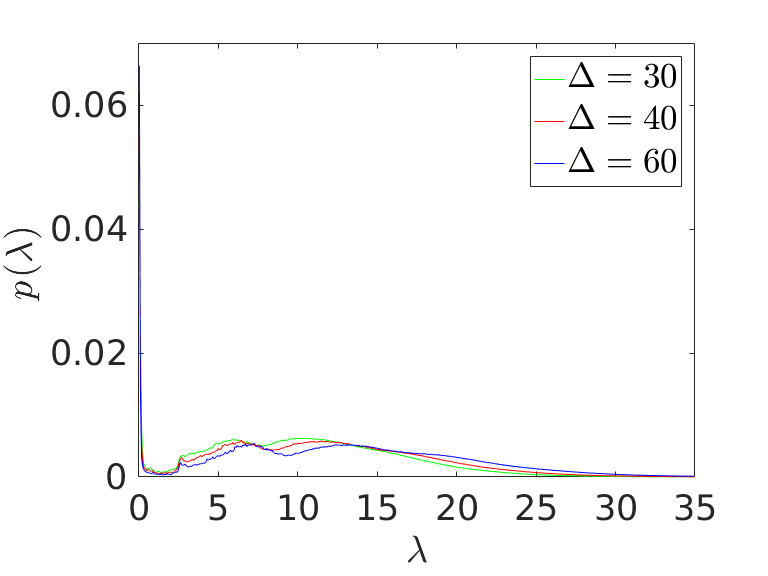}}
\put(-10,200){\textbf{b}} 
\end{picture} \\
\begin{picture}(204,200)
\put(0,0){\includegraphics[width=0.45\textwidth]{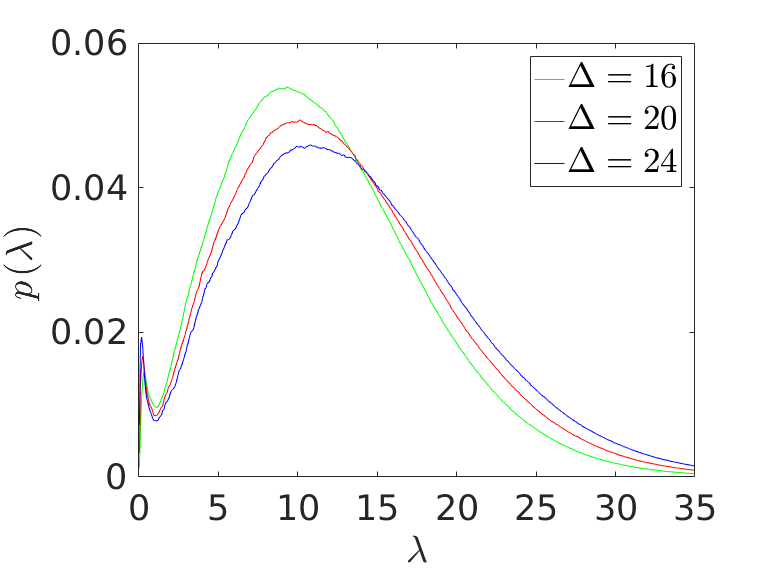}}
\put(-10,200){\textbf{c}}
\end{picture}
 \ \ \ \ \ \ \ \ \ \ 
\begin{picture}(220,200)
\put(0,0){\includegraphics[width=0.45\textwidth]{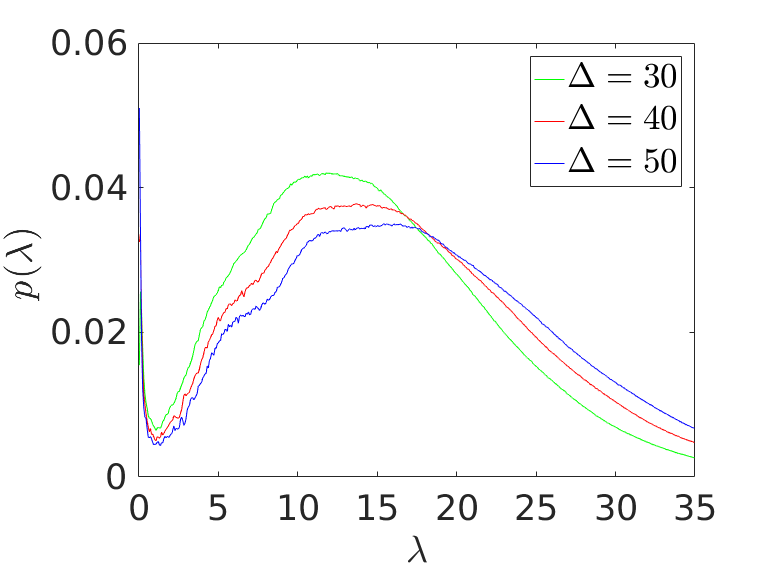}}
\put(-10,200){\textbf{d}}
\end{picture} 
  \caption{\textbf{a}, \textbf{b}: Probability density for $n_\mr{max} = 1$  of the entanglement energies of the reduced density matrices of all $2 \times 2$ plaquettes sampling over $10^4$ eigenstates for each $4 \times 4$ plaquette. The inset shows an enlargement of the distribution at low entanglement energies.
  \textbf{c}, \textbf{d}, Same probability density for $n_\mr{max} = 2$, $U' = 2$ for all $2 \times 2$ plaquettes sampling over 1,000 eigenstates for each plaquette. The bumps in the broad peak are mostly due to changes in the entanglement properties between even and odd sites (depending on the position $(x,y)$ of the plaquette, its causal cone can vary from size $4 \times 4$ to $6 \times 6$).}
\label{fig:entspec}
\end{figure*}

In Fig.~\ref{fig:entropies_U2} we present the data for the distribution of entropies and on-site occupations for $U' = 2$. Finally, we calculated the distribution of the entanglement energies for $2 \times 2$ plaquettes and $n_\mr{max} = 1$ as well as $n_\mr{max} = 2$, $U' = 2$:  
The entanglement energies are the eigenvalues $\lambda_k$ of the \textit{entanglement Hamiltonian} $H_\mr{ent}$ defined via $\rho_{2 \times 2} = \exp\left(-H_\mr{ent}\right)$. 
 The corresponding distributions are shown in Fig.~\ref{fig:entspec}. They have a double-peak structure throughout with a narrow maximum at small entanglement energies and a broad maximum at higher entanglement energies $\{ \lambda_k\}$. The former moves towards zero and the latter towards infinite entanglement energies with increasing $\Delta$. The behavior of the narrow maximum is consistent with findings using exact diagonalization in one dimension~\cite{Geraedts2017}. However, in that reference, the broad maximum is observed only in the MBL phase and close to the transition. It is also found to move to larger entanglement energies with increasing disorder strength. It is thought to be a remnant of the critical regime even deep in the localized phase. We obtain a broad maximum for small $\Delta$, too, presumably because the optimized tensor network does not fully capture the delocalized phase, and instead displays critical features there also.

\section{Transition for random potentials chosen from a uniform distribution}

\begin{figure*}
\begin{picture}(220,200)
\put(0,0){\includegraphics[width=0.45\textwidth]{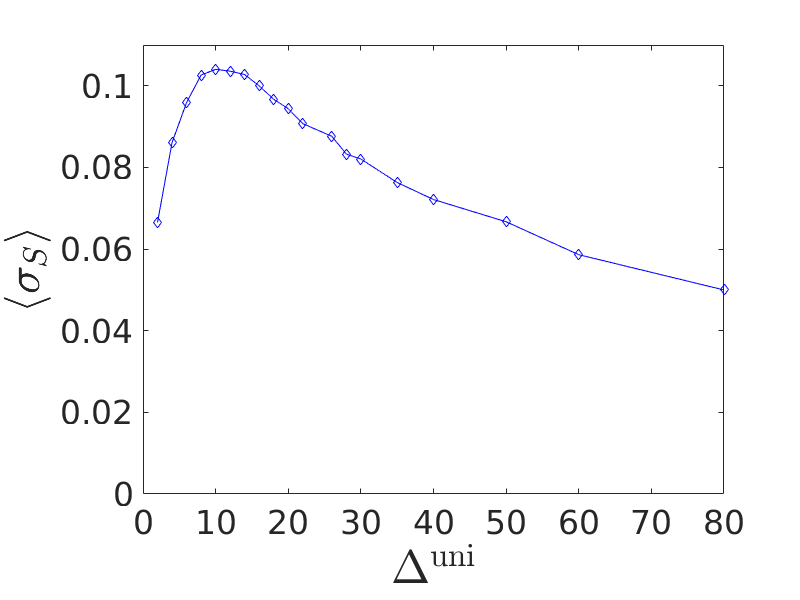}}
\put(-10,200){\textbf{a}}
\end{picture}
 \ \ \ \ \ 
\begin{picture}(220,200)
\put(0,0){\includegraphics[width=0.45\textwidth]{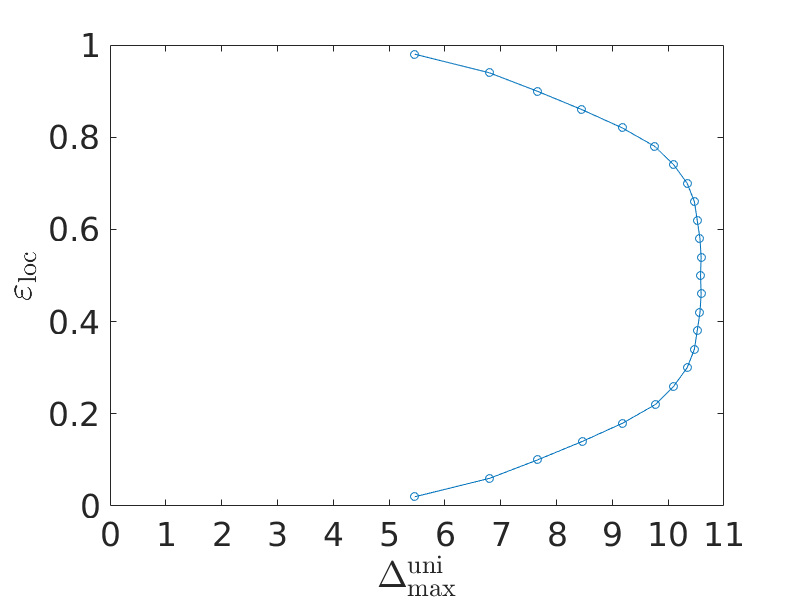}}
\put(-10,200){\textbf{b}} 
\end{picture} 
  \caption{\textbf{a}, Entropy fluctuation for the random disorder given by a uniform distribution $[-\Delta^\mr{uni},\Delta^\mr{uni}]$ and $n_\mr{max} = 1$. \textbf{b}, Corresponding mobility edge. These results were obtained in the same way as in the main text. A fit to the entropy fluctuation by a fraction of second order polynomials yields $\Delta_c^\mr{uni} = 10.5$ and the extremal point of the mobility edge $\Delta_c^\mr{uni} = 10.6$.}
\label{fig:box}
\end{figure*}

We carried out a series of tensor network optimizations for the Hamiltonian of Eq.~(3) and $n_\mr{max} = 1$ with the random disorder potential $\delta_i$ given by a uniform distribution from the interval $[-\Delta^\mr{uni},\Delta^\mr{uni}]$. We used 30 disorder realizations per disorder strength $\Delta^\mr{uni}$ for a $10 \times 10$ lattice and calculated the entropy fluctuation and mobility edge in the same way as in the main body. The results are shown in Fig.~\ref{fig:box}. They indicate a cirtical disorder strength of $\Delta_c^\mr{uni} = 10.6$. This value is smaller than what one might anticipate after acounting for the different variances: The variances are $\sigma_\mr{Gaussian} = \frac{\Delta}{2 \sqrt{2 \ln(2)}}$ and $\sigma_\mr{uniform} = \frac{\Delta^\mr{uni}}{\sqrt{3}}$. Hence, one would expect a transition at $\frac{1}{2} \sqrt{\frac{3}{2 \ln(2)}} \Delta_c = 0.736 \times 19 = 14$ for the uniform distribution. The probability density of on-site entropies has again a clear bimodal distribution near $\Delta_c^\mr{uni}$ (not shown), consistent with an MBL-to-thermal transition.

\bibliography{biblioMBL}{}

\end{document}